\documentclass[10pt, conference, letterpaper]{ieeeconf}
\IEEEoverridecommandlockouts
\usepackage{kpmp_defs}

\title{\LARGE \bf A frequency-constrained geometric Pontryagin maximum principle on matrix Lie groups%
\thanks{The authors are with Systems \& Control Engineering, IIT Bombay, Powai, Mumbai 400076, India, and acknowledge the support of the grant 17ISROC001 from the Indian Space Research Organization.}%
}

\author{%
	Shruti Kotpalliwar,
	Pradyumna Paruchuri,
	Karmvir Singh Phogat,
	Debasish Chatterjee,
	Ravi Banavar%
	\thanks{Emails: \texttt{\{shruti, pradyumn, karmvir.p, chatterjee, banavar\} @sc.iitb.ac.in}}%
}

\begin{document}

	\maketitle

	\begin{abstract}
	%
	%
		In this article we present a geometric discrete-time Pontryagin maximum principle (PMP) on matrix Lie groups that incorporates frequency constraints on the controls in addition to pointwise constraints on the states and control actions directly at the stage of the problem formulation. This PMP gives first order necessary conditions for optimality, and leads to two-point boundary value problems that may be solved by shooting techniques to arrive at optimal trajectories. We validate our theoretical results with a numerical experiment on the attitude control of a spacecraft on the Lie group $\SO(3)$.
	%
	\end{abstract}

	\section{Introduction}
		\label{sec:the intro}
		Most engineering systems are required to operate in a certain pre-defined region of the state and control spaces. For instance, since mechanical systems are inertial, mechanical actuators have natural limitations in terms of, e.g., the torque magnitudes and the operating frequencies. In the control literature these are known as control magnitude and frequency constraints, respectively. Control magnitudes must be limited, for instance, to prevent rapid movements of robotic arms for safety considerations \cite{Tokhi}. Frequency constraints arise from a more subtle consideration. Consider, for instance, read/write operations in disk drives \cite{Brockett} where excitation of the actuator at flexible modes may result in erroneous read/write operations, attitude orientation manoeuvres of satellites fitted with flexible structures such as solar panels \cite{Gorinevsky} may excite the natural frequencies of the flexible structures, leading to vibrations and structural damage unless the natural frequencies are avoided, etc. It is, therefore, desirable to eliminate certain frequencies from the spectra of the control functions of controlled systems at the control synthesis stage.

Traditional attempts by control engineers to handle frequency constraints includes filtering of the actuating signal after control synthesis, or techniques of more recent vintage such as \(H_\infty\) control \cite{Doyle92} that minimize a weighted combination of transfer functions in certain frequencies as part of the synthesis technique. Both these techniques suffer from their own problems: The former is {\it ad hoc} and based on the designer's intuition of the system and actuator, and the latter, though more systematic and incorporates penalties on the frequencies in an interval, still suffers from the inability to completely suppress a pre-specified set of frequencies. More importantly, none of these techniques is capable of incorporating hard bounds on the control actions and the states. Constraints on the control actions and the states are present in the classical optimal control paradigm \cite{ref:Bol-75}, but frequency constraints in the same framework have not been treated in the literature so far, with the exception of \cite{PradPMP}, where we presented a set of first order necessary conditions for optimal control problems with frequency constraints on the control action trajectories. This article is a continuation of our studies to more applied problems for an important class of mechanical systems.

The configuration variables of a large class of mechanical systems (e.g., spacecraft, mobile robots, autonomous underwater vehicles \cite{Wadoo2017},) evolve on matrix Lie groups. Developing a control paradigm for such a class of systems, where hard constraints on the actuators and the states, as well as frequency constraints on the control actions, would prove invaluable to the community of control practitioners. The problem described above can be stated as a constrained optimal control problem, and optimal control theoretic \cite{ref:Bol-75} tools such as the Pontryagin maximum principle (PMP) and dynamic programming can, in principle, be applied to them. 

Optimal control problems for controlled mechanical systems evolving on non-flat manifolds cannot be solved using the discrete-time PMP on Euclidean spaces \cite{ref:Bol-75} because this technique does not carry over directly to such manifolds. We developed a geometric version of the PMP for constrained optimal control problems on matrix Lie groups in \cite{KarmvirPMP}, but frequency constraints on the control actions were not considered there. A comprehensive framework for constrained discrete-time optimal control problems on matrix Lie groups with state-action constraints and frequency constraints on the control actions is needed, and in this article we establish a discrete-time geometric version of the PMP tailored to such problems. The main contribution of this article is that the technique presented here provides \emph{tractable} solutions to optimal control problems, incorporating, at once, an entire class of hard constraints on the states, control actions, and frequency constraints on the control trajectories, right at the synthesis stage. The constrained two point boundary value problems arising out of the necessary conditions for optimality can be solved via multiple shooting techniques that can be implemented on a parallel architecture for fast computation. 

After introducing the necessary notations, we expose the precise problem statement in \secref{sec:the prob}, and follow up with the main result --- a geometric discrete-time PMP on matrix Lie groups --- in \secref{sec:main result}. A proof of the main result is given in \secref{sec:proof}, and numerical experiments on spacecraft attitude control are presented in \secref{sec:numerical simulations}.

	\section{Problem setup}
		\label{sec:the prob}
Fix a positive integer \(\horizon\) that will play the r\^ole of a time horizon. For \(\mathcal{N} \in \N\), we set \(\Scale[\mathcal{N}] \Let \set{0, \ldots, \mathcal{N}-1}\) and \(\scaleZ[\mathcal{N}] \Let \Scale[\mathcal{N}] \setminus \set{0}\). Let \(\LieGroup\) be a matrix Lie group with \(\LieAlg\) its Lie algebra; the map \(\expMap \! : \LieAlg \ra \LieGroup\) denotes the corresponding exponential map. Recall  \cite[p.\ 273]{Marsden2013} that for each $X \in \LieAlg$, there exist a map $\expMap^{X}(\cdot):\R \rightarrow \LieGroup$ such that $\expMap^X(0) = e \in \LieGroup$, \(\partial_t|_{t=0} \expMap^{X}(t) = X,\) and \(\expMap^X (t+s)=\expMap^X (t)\expMap^X (s)\), where \(e\) is the group identity. For \(y\in \C^{\horizon}\) we let \( \support (y) \Let \set{ i \in \scaleZ[\horizon] \suchthat y_{i} \neq 0}. \)

Consider a controlled discrete-time system evolving partly on a fixed matrix Lie group \(G\) and partly on \(\R^d\), given by
\begin{equation}
\label{e:system}
\begin{aligned}
\begin{cases}
	\config_{t+1} = \config_{t} \intStep (\config_{t}, \state_{t})\\
	\state_{t+1} = \sysDyn (\config_{t}, \state_{t}, \conInp_{t})
\end{cases} & \text{for } t \in \scaleZ,
\end{aligned} \end{equation}
where \((\config_{t}, \state_{t}) \in \LieGroup \times \R^{\sysDim}\) is the vector of states and \(\conInp_{t} \in \R^{\conDim}\) the vector of control actions of the system at a discrete-time instant \(t\). The maps \(\intStep \! : \LieGroup \times \R^{\sysDim} \ra \LieGroup\) describing the dynamics of the states \(\config_{t}\) on the matrix Lie group \(\LieGroup\), and \(\sysDyn \! : \LieGroup \times \R^{\sysDim} \times \R^{\conDim} \ra \R^{\sysDim}\) describing the dynamics of state \(\state_{t}\) in \(\R^{\sysDim}\) are smooth. We further assume that there exists an open set \(\openSetAlg \subset \LieAlg\) such that the following conditions hold:
\begin{enumerate}[label=(\alph*), leftmargin=*, align=left, widest=a,]
	\item \label{diffeo} the exponential map \(\expMap \! : \openSetAlg \ra \expMap (\openSetAlg)\) is a diffeomorphism, and 
	\item \label{uniformity} the image of \(\intStep\) is a subset of \(\expMap (\openSetAlg)\) for all \(t\).
\end{enumerate}
	An open set \(\openSetAlg\) satisfying the condition \ref{diffeo} always exists by definition of the exponential map; the condition \ref{uniformity} is, however, an \emph{assumption}, in effect stipulating that the time-discretization step is sufficiently small.

\subsubsection*{Frequency constraints}
	We provide a brief discussion on discrete-time control frequency constraints. Let \(\R^{\horizon} \ni \conInp\kth[k] \Let ( \conInp_{t}\kth[k] )_{t=0}^{\horizon-1} \) be the trajectory of the \(k^{\text{th}}\) component of the control.
Throughout the article, the subscript on $\conInp$ denotes the stage and the superscript denotes the component of the control. The hat on top of a variable denotes its frequency representation. The discrete Fourier transform (DFT) of \(\conInp\kth[k]\) is defined by \cite[Chapter 7]{Stein-Shakarchi}
\[
	\mathbb{C}^{\horizon} \ni \freqComp \Let \bm{\dft}\conInp \kth \quad \text{ for } k = 1, \ldots, \conDim,
\]
where 
\(\bm{\dft}\Let \frac{1}{\sqrt{\horizon}} 
\begin{pmatrix}
1& 1 & \ldots & 1 \\
1& \omega & \ldots & \omega^{\horizon-1}\\
\vdots & \vdots & \ddots & \vdots \\
1 & \omega^{\horizon-1} & \ldots & \omega^{(\horizon-1)(\horizon-1)}
\end{pmatrix} \in \mathbb{C}^{\horizon \times \horizon}\)
for \(\omega \Let \epower{\frac{- i 2\pi }{\horizon}} \). 
%
%
	We let \(\conInp\) denote the stacked vector \(\pmat{{(\conInp \kth[1])} \transpose & \ldots & {(\conInp\kth[\conDim])} \transpose} \transpose\), and define the DFT of a control trajectory by the vector 
\[
\C^{\conDim \horizon} \ni \fullFreqComp \Let 
\begin{pmatrix}
  \freqComp[1]\\ \vdots \\ \freqComp[\conDim] 
\end{pmatrix} 
=\begin{pmatrix}
\bm{\dft} \conInp \kth[1]\\ \vdots \\ \bm{\dft}\conInp \kth[m]
\end{pmatrix} 
=\diagDFT 
\begin{pmatrix}
 \conInp \kth[1]\\ \vdots \\ \conInp \kth[m]
\end{pmatrix},
\] where \(\diagDFT\) is a block diagonal matrix with the standard DFT matrix \( \bm{\dft} \) being each block. Note that \((\freqComp)_{j} \in \C \) represents the \((2\pi(j-1)/N)^{\text{th}}\) frequency component of the trajectory \(u\kth\). 
%
%
%
Therefore, if elimination of the \((2\pi(j-1)/N)^{\text{th}} \) frequency component of \(u\kth \) is desired, it is ensured by introducing the constraint
\[  
   0 = (\freqComp)_{j} = \bm{\dft}_j \conInp \kth,
\]
where \(\bm{\dft}_j\) is the \(j^{\text{th}}\) row of the DFT matrix defined above. Therefore, in general, control frequency constraints are enforced by a collection of affine equality conditions in the control action variables, and we represent them in an abstract fashion by one equality constraint
\begin{equation}
\label{eq:Frequency Constraints}
 \sum_{t=0}^{\horizon-1} \freqDer \conInp_{t} =0 
 \quad\text{where \(\freqDer\) are suitable matrices}.
\end{equation}
The reader will notice that the manner in which frequency constraints are assimilated into the problem formulation enables the designer to cancel particular frequencies in the control inputs, a feature distinctly absent in other control synthesis schemes. (A more detailed discussion on control frequency constraints may be found in \cite{PradPMP}.) 

Collecting the definitions above, we write our constrained optimal control problem in discrete-time:
\begin{gather}
	\label{e:the problem}
	\begin{aligned}
		& \minimize_{(\conInp_{t})_{t=0}^{\horizon-1}} && \sum_{t=0}^{\horizon-1} \cost (\config_{t}, \state_{t}, \conInp_{t}) + \cost[\horizon] (\config_{\horizon}, \state_{\horizon})\\
		& \sbjto && \begin{cases}
						\text{dynamics \eqref{e:system}}, \\
						\conInp_{t} \in \admControl \quad \text{for each } t \in \Scale, \\
						\stateConstr (\config_{t}, \state_{t}) \le 0 \quad \text{for each } t \in \scaleZ[\horizon+1]\\
						(\config_{0}, \state_{0}) = (\cnfgPivot, \stPivot),\\ 
						\freqConstr (\conInp_{0}, \ldots, \conInp_{\horizon-1}) = 0,
					\end{cases}
	\end{aligned} \raisetag{4\baselineskip}
\end{gather}
with the following data:
\begin{enumerate}[label=(\ref{e:the problem}-\roman*), leftmargin=*, widest=iii, align=left]
	\item \((\cnfgPivot, \stPivot) \in \LieGroup \times \R^{\sysDim}\) and \(\horizon \in \Nz \) are fixed; 
	\item the maps \(\cost \! : \LieGroup \times \R^{\sysDim} \times \R^{\conDim} \ra \R\), for each \(t \in \Scale\) defining cost-per-stage and \(\cost[\horizon] \! : \LieGroup \times \R^{\sysDim} \ra \R\) accounting for the final stage cost are smooth;
	\item the maps \(\stateConstr \! : \LieGroup \times \R^{\sysDim} \ra \R^{\stConDim}\) for \(t \in \scaleZ[\horizon+1] \) denote constraints on the states and are smooth;
	\item the set of admissible control actions \(\admControl \subset \R^{\conDim}\) is convex and compact for each \(t \in \Scale \);
	\item \label{freq map} the linear map \(\R^{\conDim \horizon} \ni (\conInp_{0}, \ldots, \conInp_{\horizon-1}) \mapsto \freqConstr (\conInp_{0}, \ldots, \conInp_{\horizon-1}) \Let \sum_{t=0}^{\horizon-1} \freqDer \conInp_{t} \in \R^{\freqDim}\) represents constraints on the frequency components of the control profile \((\conInp_{t})_{t=0}^{\horizon-1}\).
\end{enumerate}

\remark{%
		The map \(\freqConstr\) defined in \ref{freq map} provides the real and imaginary components of the required frequency components of the control profile. In this article the main result is specialized to eliminating certain frequency components and hence we simply set the constraint \(\freqConstr (\conInp_{0}, \ldots, \conInp_{\horizon-1}) = 0\) in \eqref{e:the problem}. However, this approach can be extended to a larger class of constraints on the frequency components; see Remark \ref{rm:freq bounds} for modifications to the approach in such cases.
		}

To state our main result, we need a few definitions from the theory of Lie groups that are relevant to this article; a detailed discussion may be found in \cite[p.\ 124, 173, 311]{Marsden2013}.

	Let \(\LieGroup \ni q \mapsto h(q) \in \R \) be a function defined on a manifold \(\LieGroup\). The \emph{tangent lift} of the function \(h\) at a point \(q_{0}\in \LieGroup\) is the map
	\[
		T_{q_{0}}\LieGroup \ni v \mapsto \derivative{q} h(q_{0})v \Let \partial_t\big|_{t=0} h(\gamma(t)) \in \R,
	\]
	where \(\gamma(t)\) is a path in the manifold \(\LieGroup\) with \(\gamma(0)=q_0\) and \(\partial_t\big|_{t=0} \gamma(t)=v.\) Let \(\Action[] \! : \LieGroup \times \LieGroup \ra \LieGroup\) be a \emph{left action} and let \(\Action \! : \LieGroup \ra \LieGroup\) denote \(\Action[](g, \cdot)\).\footnote{The \emph{left action} on a Lie group should not be confused with control \emph{actions} in the context of our control system.} The \emph{tangent lift} of \(\Action[]\), \(\tanLift{\Action[]} \! : \LieGroup \times \tanLift{\LieGroup}\) is the action
	\[
		(\elem, (h, v)) \mapsto \tanLift{\Action} (h, v) = \bigl(\Action (h), \tanLift[h]{\Action} (v) \bigr), \quad  (h, v) \in \tanLift[h]{\LieGroup}.
	\]
	The \emph{cotangent lift} of \(\Action[]\), \(\cotanLift{\Action[]} \! : \LieGroup \times \cotanLift{\LieGroup} \ra \cotanLift{\LieGroup}\), is the action
	\[
		(\elem, (h, a)) \mapsto \cotanLift{\Action[\elem \inverse]} (h, a) = \bigl(\Action (h), \cotanLift[\Action (h)]{\Action[\elem \inverse]} (a) \bigr), \; a \in \cotanLift[h]{\LieGroup}.
	\]
	The \emph{adjoint action} of \(\LieGroup\) on \(\LieAlg\) is defined as
	\[
		\LieGroup \times \LieAlg \ni (\elem, \beta) \mapsto \adjAction_{\elem} \beta \Let \partial_s\bigr|_{s=0} \elem \epower{s \beta} \elem \inverse \in \LieAlg,
	\]
	and finally, the \emph{co-adjoint action} of \(\LieGroup\) on \(\dualSpace{\LieAlg}\) is the inverse dual of the adjoint action, given by
	\[
		\LieGroup \times \dualSpace{\LieAlg} \ni (\elem, a) \mapsto \adjAction_{\elem \inverse}^{*} a \in \dualSpace{\LieAlg},
	\]
	where \(\inprod{\adjAction_{{\elem \inverse}}^{*} a}{\beta} = \inprod{a}{\adjAction_{\elem \inverse} \beta}\).

	\section{Main result}
		\label{sec:main result}
		The following theorem is our main result:
\begin{theorem}
	\label{th:main}
	Let \(\bigl(\optInp \bigr)_{t=0}^{\horizon-1}\) be an optimal control trajectory for \eqref{e:the problem} and \(\bigl(\optConfig, \optState \bigr)_{t=0}^{\horizon}\) be the corresponding state trajectory. Define the Hamiltonian
	\begin{gather}
		\label{e:Hamiltonian}
		\begin{aligned}
			& \LieAlg \times \R^{\sysDim} \times \scaleZ \times \LieGroup \times \R^{\sysDim} \times \R^{\conDim} \ni \bigl(\genGroup, \genDyn, \genTime, \genConfig, \genState, \genInp \bigr) \mapsto \\
			& \Hamiltonian[\genCost, \genFreq] (\genGroup, \genDyn, \genTime, \genConfig, \genState, \genInp) \Let \genCost \cost[\genTime](\genConfig, \genState, \genInp) + \inprod{\genGroup}{\expMap \inverse \bigl(\intStep[\genTime](\genConfig, \genState) \bigr)}_{\LieAlg} \\
			& \qquad \qquad \qquad \qquad \quad + \inprod{\genDyn}{\sysDyn[\genTime](\genConfig, \genState, \genInp)} + \inprod{\genFreq}{\freqDer[\genTime] \genInp} \in \R.
		\end{aligned} \raisetag{3.2\baselineskip} 
	\end{gather}
	Then there exist
	\begin{itemize}[label=\(\circ\), leftmargin=*]
		\item adjoint trajectories \(\bigl(\adjDyn \bigr)_{t=0}^{\horizon-1} \subset \R^{\sysDim} , \bigl(\adjGroup \bigr)_{t=0}^{\horizon-1} \subset \LieAlg\),
		\item covectors \(\adjState \in \R^{\stConDim} \) for \(t \in \scaleZ[\horizon+1]\), and
		\item a pair \(\bigl(\adjCost, \adjFreq \bigr) \in \set[\big]{-1, 0} \times \R^{\freqDim} \)
	\end{itemize}
	such that, with \(\extremal \Let \bigl(\adjGroup, \adjDyn,  t,\optConfig, \optState, \optInp \bigr)\) and \(\DummyadjGroup \Let \codexp\bigl(\adjGroup \bigr) \), the following conditions hold:
	\begin{enumerate}[label={\textup{(\roman*)}}, leftmargin=*, align=left, widest=iii]
		\item \label{pmp:non-triviality} the non-triviality condition:
			\begin{quote}
				the adjoint variables \(\bigl(\adjDyn \bigr)_{t=0}^{\horizon-1}\), \(\bigl(\adjGroup \bigr)_{t=0}^{\horizon-1}\), and the pair \((\adjCost,\; \adjFreq)\) do not simultaneously vanish;
			\end{quote}
		\item \label{pmp:dynamics} the state and adjoint dynamics:
			\begin{align*}
				\text{state} \quad & \begin{cases}
										\ \optConfig[t+1] = \optConfig[t] \exp \bigl(\derivative{\genGroup} \Hamiltonian (\extremal) \bigr), \\
										\ \optState[t+1] = \derivative{\genDyn} \Hamiltonian (\extremal),
									 \end{cases}\\
				\text{adjoint} & 	\begin{cases}
										\ \DummyadjGroup[t-1] = \triv{\optConfig} \Bigl(\derivative{\genConfig} \Hamiltonian(\extremal) + \adjState \derivative{\genConfig} \stateConstr (\optConfig, \optState) \Bigr)\\
										\qquad\qquad + \ \adjAction_{\exp \bigl(-\derivative{\genGroup} \Hamiltonian (\extremal)\bigr)}^{*} \DummyadjGroup,\\
										\ \adjDyn[t-1] = \derivative{\genState} \Hamiltonian (\extremal) + \adjState \derivative{\genState} \stateConstr (\optConfig, \optState);
									\end{cases}
			\end{align*}
		\item \label{pmp:transversality} the transversality conditions:
			\begin{align*}
				& \DummyadjGroup[\horizon-1] = \triv{\optConfig[\horizon]} \Bigl(\adjCost \derivative{\genConfig} \cost[\horizon] (\optConfig[\horizon], \optState[\horizon]) + \adjState[\horizon] \derivative{\genConfig} \stateConstr[\horizon](\optConfig[\horizon], \optState[\horizon]) \Bigr),\\
				& \adjDyn[\horizon-1] = \adjCost \derivative{\genState} \cost[\horizon] (\optConfig[\horizon], \optState[\horizon]) + \adjState[\horizon] \derivative{\genState} \stateConstr[\horizon] (\optConfig[\horizon], \optState[\horizon]);
			\end{align*}
		\item \label{pmp:ham maximization} the Hamiltonian non-positive gradient condition:
			\[
				\inprod{\derivative{\genInp} \Hamiltonian(\extremal)}{\genVec - \optInp} \le 0 \quad \text{for all } \genVec \in \admControl;
			\]
		\item \label{pmp:slackness} the complementary slackness conditions:
			\begin{align*}
				{(\adjState)}^{(j)} &\stateConstr^{(j)} (\optConfig, \optState) = 0  \\ 
				&\text{for all } j \in \scaleZ[\stConDim+1] \text{ and } t \in \scaleZ[\horizon+1] 
			\end{align*}
		\item \label{pmp:non-positivity} the non-positivity condition
			\[
				\adjState \le 0 \quad \text{for all } t \in \scaleZ[\horizon+1].
			\]
	\end{enumerate}
\end{theorem}

 \remark{
	The adjoint variables (a.k.a.\ `multipliers',) corresponding to the cost, the dynamics, the state-constraints, and the frequency constraints of the control trajectories appear here, and we distinguish between them by introducing the different super-scripts of the single Greek letter \(\eta\). Various objects in frequency space are distinguished by a `hat'. In particular, the two adjoint variables that are constant with time appear in the superscript of the Hamiltonian.
	}

	\section{Proof of the main result}
		\label{sec:proof}
		In this section we first provide a sketch of a proof of Theorem \ref{th:main}, and subsequently elaborate on each step of the proof.
\begin{enumerate}[label=(S-\roman*), leftmargin=*, widest=iii, align=left]
	\item \label{pf:step1} The frequency constraints are represented as an auxiliary dynamical system that is incorporated into the optimal control problem.
	\item \label{pf:step2} We define a diffeomorphism to translate the optimal control problem obtained in step \ref{pf:step1} to an equivalent optimal control problem on a Euclidean space of appropriate dimension;
	\item \label{pf:step3} The optimal control problem obtained in \ref{pf:step2} is converted to a static optimization problem, and first order necessary conditions for optimality are derived using Boltyanskii's method of tents \cite{ref:Bol-75}.
	\item \label{pf:step4} The first order necessary conditions are mapped from the Euclidean space to the configuration space via the cotangent lift of the diffeomorphism.
\end{enumerate}

\subsubsection*{\ref{pf:step1}: Frequency constraints via a dynamical system}
	The frequency constraints in \eqref{e:the problem} are defined via the linear maps \(\freqDer: \R^{\conDim} \to \R^{\freqDim} \) defined in \eqref{eq:Frequency Constraints}. We recast these constraints in the form of a linear controlled dynamical system in an auxiliary variable $w$ as follows:
	\begin{equation}
		\label{e:aux sys}
		 \auxState[t+1] = \auxState[t] + \freqDer \conInp_{t}, \quad \auxState[0] = 0 \in \R^{\freqDim}. 
	\end{equation}
	To wit, the frequency constraints in \eqref{e:the problem} are defined by the linear dynamics \eqref{e:aux sys} together with the boundary condition \(\auxState[\horizon] = 0\). Therefore, replacing the frequency constraints with the linear dynamics \eqref{e:aux sys}, we write \eqref{e:the problem} in a standard form:
	\begin{gather}
		\label{e:mod problem}
		\begin{aligned}
			& \minimize_{(\conInp_{t})_{t=0}^{\horizon-1}} && \sum_{t=0}^{\horizon-1} \cost (\config_{t}, \state_{t}, \conInp_{t}) + \cost[\horizon] (\config_{\horizon}, \state_{\horizon})\\
			& \sbjto && \begin{cases}
							\text{dynamics \eqref{e:system}\text{ and }\eqref{e:aux sys}},\\
							\conInp_{t} \in \admControl \quad \text{for each } t \in \Scale,\\
							\stateConstr (\config_{t}, \state_{t}) \le 0 \quad \text{for each } t \in \scaleZ[\horizon+1],\\
							(\config_{0}, \state_{0}, \auxState[0]) = (\cnfgPivot, \stPivot, 0), \ \text{and } \auxState[\horizon] = 0.		
						\end{cases}
		\end{aligned} \raisetag{3\baselineskip}
	\end{gather}

	\remark{\label{rm:freq bounds} With the auxiliary system defined as in (5), the final states \(\auxState[\horizon]\) correspond to the real and imaginary parts of the required/chosen frequency components. Therefore constraints on the frequency components of the control profile translate to final state constraints on the auxiliary states.}

\subsubsection*{\ref{pf:step2}: Translation of \eqref{e:mod problem} to Euclidean space}
	Let us define a parametrization of the Lie group \( \LieGroup\) in an open neighbourhood of \(\config_{t}\) for each \(t \in \Scale[\horizon+1] \). Define an open neighborhood of \(\config_t\) as \(\mathcal{Q}_{t} \Let \set{\Action[\config_{t}] (\elem) \suchthat \elem \in \exp (\openSetAlg)}\), where \(\openSetAlg\) is an open set in the Lie algebra \(\LieAlg\) containing \(0\). Then for a given \(\config_{t} \in \LieGroup\), the map \(\action{\config_{t}} \Let \bigl(\Action[\config_{t}] \circ \exp \bigr) \inverse :\mathcal{Q}_{t} \ra \openSetAlg \subset \LieAlg\) lends a unique representative for \(\config_{t+1} \in \LieGroup\) on the Lie algebra \(\LieAlg\) for all \(t \in \Scale\).
	
	 Since \(\LieGroup \) is a matrix Lie group, the corresponding Lie algebra \(\LieAlg\) is a finite dimensional vector space, and hence there exists a linear homeomorphism \(\algToEucl \colon \LieAlg \ra \R^{\algDim}\), where \(\algDim\) is the dimension of Lie algebra \(\LieAlg\). The linear homeomorphism \(\algToEucl\) further translates the dynamics from the Lie algebra to an Euclidean space. We
	 now provide a detailed description of the translation of the optimal control problem to the Euclidean space. 
	For the sake of brevity, define \(\configSeq \Let (\config_{0}, \ldots, \config_{\horizon}), \quad \stateSeq \Let(\state_{0}, \ldots, \state_{\horizon}), \auxStateSeq \Let (\auxState[0], \ldots, \auxState[\horizon]), \quad \inputSeq \Let (\conInp_{0}, \ldots, \conInp_{\horizon-1})\),
	and a product manifold 
	\[
		\prodManifold \Let \LieGroup^{\horizon+1} \times \R^{\sysDim (\horizon + 1)} \times \R^{\freqDim (\horizon + 1)} \times \R^{\conDim \horizon},
	\]
	such that every state and action trajectory corresponds to a unique point on \(\prodManifold\), i.e., \( (\configSeq, \stateSeq, \auxStateSeq, \inputSeq)  \in \prodManifold. \)
			
	We define a map from the open set 
	\begin{align*}
		\EuclSpace & \Let \algToEucl(\openSetAlg)^{\horizon + 1} \times \R^{(\horizon + 1) \sysDim} \times  \R^{(\horizon + 1) \freqDim} \times \R^{\horizon \conDim}\\
		& \subset \R^{(\horizon + 1) \algDim} \times \R^{(\horizon + 1) \sysDim} \times \R^{(\horizon + 1) \freqDim} \times \R^{\horizon \conDim}
	\end{align*}
	into an open subset of \(\prodManifold\) that enables us to translate the optimal control problem \eqref{e:mod problem} to a Euclidean space as
	\begin{multline*}
		\EuclSpace \ni (\dummyConfigEuclSeq, \stateSeq, \auxStateSeq, \inputSeq) \mapsto \proTransfer (\dummyConfigEuclSeq, \stateSeq, \auxStateSeq, \inputSeq)\\
		\Let \bigl(\EuclToGroup[0] (\dummyConfigEuclSeq), \ldots, \EuclToGroup[\horizon] (\dummyConfigEuclSeq), \stateSeq, \auxStateSeq, \inputSeq) \bigr) \in \proTransfer (\EuclSpace),
	\end{multline*}
	where for \(t \in \Scale[\horizon + 1]\) and a fixed \(\cnfgPivot \in \LieGroup\),
	\begin{equation}
	\label{e: Eucl To Group map}
	\EuclToGroup (\dummyConfigEuclSeq) \Let \cnfgPivot \exp \bigl(\algToEucl \inverse (\dummyConfigEucl[0]) \bigr) \cdots \exp \bigl(\algToEucl \inverse (\dummyConfigEucl) \bigr).
	\end{equation}
	Observe that the map \(\proTransfer\) is a smooth bijection whose inverse is given by the smooth map
	\begin{align*}
		&\proTransfer (\EuclSpace) \ni (\dummyConfigSeq, \stateSeq, \auxStateSeq, \inputSeq) \mapsto \proTransfer \inverse (\dummyConfigSeq, \stateSeq, \auxStateSeq, \inputSeq)\\
		& \Let \biggl(\bigl(\algToEucl \circ \exp \inverse \bigr) (\cnfgPivot \inverse \dummyConfig[0]), \ldots, \bigl(\algToEucl \circ \exp \inverse \bigr) (\dummyConfig[\horizon-1] \inverse \dummyConfig[\horizon]), \stateSeq, \auxStateSeq, \inputSeq \biggr). 
	\end{align*}
	In other words, \(\proTransfer\) is a diffeomorphism. It is important to note that all the feasible state-action trajectories of \eqref{e:mod problem} lie in the image of \(\proTransfer\) as discussed in \cite{KarmvirPMP}.
	We employ the diffeomorphism \(\proTransfer\) to translate \eqref{e:mod problem} from the manifold \(\prodManifold\) to the Euclidean space, and for \(\EuclConfigSeq \Let (\EuclConfig_{0}, \ldots, \EuclConfig_{\horizon}) \in \R^{(\horizon + 1) \algDim}\), we arrive at
	\begin{gather}
		\label{e:Eucl problem}
		\begin{aligned}
			& \minimize_{(\conInp_{t})_{t=0}^{\horizon-1}} && \!\! \sum_{t=0}^{\horizon-1} \cost \bigl(\EuclToGroup (\EuclConfigSeq), \state_{t}, \conInp_{t} \bigr) + \cost[\horizon] \bigl(\EuclToGroup[\horizon] (\EuclConfigSeq), \state_{\horizon} \bigr)\\
			& \sbjto &&\!\! \begin{cases}
							\begin{cases}
								\EuclConfig_{t+1} = \bigl(\algToEucl \circ \exp \inverse \circ \intStep \bigr) \bigl(\EuclToGroup (\EuclConfigSeq), \state_{t} \bigr)\\
								\state_{t+1} = \sysDyn \bigl(\EuclToGroup (\EuclConfigSeq), \state_{t}, \conInp_{t} \bigr)\\
								\auxState[t+1] = \auxState + \freqDer \conInp_{t}
							\end{cases} \! \!\!\!\!\! \text{for } t \in \Scale[\horizon],\\
							\stateConstr \bigl(\EuclToGroup (\EuclConfigSeq), \state_{t} \bigr) \le 0 \quad \text{for } t \in \scaleZ[\horizon + 1],\\
							\conInp_{t} \in \admControl \quad \text{for } t \in \Scale[\horizon],\\
							(\EuclConfig_{0}, \state_{0}, \auxState[0]) = (0, \stPivot, 0), \quad \text{and} \quad 	\auxState[\horizon]=0.
						\end{cases}
		\end{aligned}\raisetag{3\baselineskip}
	\end{gather}
	
\subsubsection*{\ref{pf:step3}: From optimal control to optimization}
	Although the optimal control problem \eqref{e:Eucl problem} is defined on a Euclidean space, the standard discrete-time PMP cannot be applied directly since the map \(\EuclToGroup\) appearing on the RHS of the first two equations leads to memory in the dynamics, i.e., \(\EuclToGroup\) depends on not just the current 
values of the states and control, but on the previous values as well. To circumvent this, we lift the optimal control problem to a static optimization problem in a higher-dimensional Euclidean space and apply Boltyanskii's method of tents \cite{ref:Bol-75} to this lifted optimization problem.

	To this end, let \(\process \Let (\EuclConfigSeq, \stateSeq, \auxStateSeq, \inputSeq) \in \R^{\proDim}\), where \(\proDim \Let (\algDim + \sysDim + \freqDim)(\horizon + 1) + \conDim \horizon\) be the stacked vector of states and control corresponding to \eqref{e:Eucl problem}. We define projection maps that allow us to access each component of \(\process\) as follows:
	\begin{gather}
		\label{e:projections}
		\begin{aligned}
			& \proj{\EuclConfig} (\process) \Let \EuclConfig_{t}, \ \proj{\state} (\process) \Let \state_{t}, \ \proj{\auxState[]} (\process) \Let \auxState \ \text{ for } t \in \Scale[\horizon + 1],\\
			& \proj{\conInp} (\process) \Let \conInp_{t} \ \text{ for } t \in \Scale \quad \text{and} \quad \proj[]{\EuclConfig} (\process) \Let \EuclConfigSeq. 
		\end{aligned} \raisetag{0.8\baselineskip}
	\end{gather}
	We lift the cost function and the constraints of \eqref{e:Eucl problem} to \(\R^{\genDim}\) using the projection maps \eqref{e:projections} as follows:
	\begin{itemize}[label=\textbullet, leftmargin=*, align=left]
		\item The condition \(\conInp_{t} \in \admControl\) becomes \(\proCon \Let \set[\big]{\process \in \R^{\proDim} \suchthat \proj{\conInp} (\process) \in \admControl}\).
		\item The end-point constraints on the states are described by \(\constrSet{0}{} \Let \set[\big]{\process \in \R^{\proDim} \suchthat \proj[0]{\EuclConfig} (\process) = 0, \ \proj[0]{\state} (\process) = \stPivot, \ \proj[0]{\auxState[]} (\process) = 0}\) and \(\constrSet{\horizon}{} \Let \set[\big]{\process \in \R^{\proDim} \suchthat \proj[\horizon]{\auxState[]} (\process) = 0}\).
		\item The cost functions, the dynamics and the state constraints are described by:
			\begin{align*}
				& \proCost (\process) \Let \sum_{t=0}^{\horizon-1} \cost \bigl(\EuclToGroup(\proj[]{\EuclConfig} (\process), \proj{\state} (\process), \proj{\conInp} (\process) )\bigr) \\
				& \phantom{\proCost (\process) \ \  \cost \bigl(\EuclToGroup \bigr)}+ \cost[\horizon] \bigl(\EuclToGroup[\horizon](\proj[]{\EuclConfig}) (\process), \proj[\horizon]{\state} (\process) \bigr) \\
				& \proAlgDyn (\process) \Let - \proj[t+1]{\EuclConfig} (\process) + \bigl(\algToEucl \circ \exp \inverse \bigr) \circ \intStep \bigl(\EuclToGroup (\proj[]{\EuclConfig} (\process), \proj{\state} (\process))\bigr) \\
				& \proStateDyn (\process) \Let - \proj[t+1]{\state} (\process) + \sysDyn \bigl(\EuclToGroup (\proj[]{\EuclConfig} (\process), \proj{\state} (\process), \proj{\conInp} (\process)) \bigr)\\
				& \proAuxDyn (\process) \Let - \auxProj[t+1] + \auxProj + \freqDer \conProj \\
				& \proStCon (\process) \Let \stateConstr \bigl(\EuclToGroup (\algSeqProj), \stProj \bigr)
			\end{align*}
	\end{itemize}

	We now arrive at a static optimization problem equivalent to the optimal control problem \eqref{e:mod problem}:
	\begin{equation}
		\label{e:opt prob}
		\begin{aligned}
			& \minimize_{\process} && \proCost (\process)\\
			& \sbjto && \begin{cases}
							\begin{cases}
								\proAlgDyn (\process) = 0\\
								\proStateDyn (\process) = 0\\
								\proAuxDyn (\process) = 0\\
								\proStCon[t+1] (\process) \le 0\\
							\end{cases} \quad \text{for } t \in \Scale[\horizon]\\
							\process \in \bigl(\bigcap_{t=0}^{\horizon-1} \proCon \bigr) \cap \constrSet{0}{} \cap \constrSet{\horizon}{}.
						\end{cases}
		\end{aligned}
	\end{equation}

	Suppose that \((\optConfigSeq, \optStateSeq, \optAuxStateSeq, \optControlSeq)\) is an optimal state-action trajectory for the optimal control problem \eqref{e:mod problem}. Then \(\optProc \Let \proTransfer^{-1}(\optConfigSeq,\optStateSeq,\optAuxStateSeq, \optControlSeq)\) is a solution of above optimization problem. By \cite[Theorem 18]{ref:Bol-75}, there exist multipliers (row vectors) \(\Bigl(\adjCost, (\dualGroup)_{t=0}^{\horizon-1},(\adjDyn)_{t=0}^{\horizon-1}, \big(\adjAuxState)_{t=0}^{\horizon-1},(\adjState)_{t=1}^{\horizon} \Bigr) \in \R \times \R^{\algDim (\horizon+1)} \times \R^{\sysDim (\horizon+1)} \times \R^{\freqDim (\horizon+1))} \times \R^{\stConDim (\horizon+1)}\), not all simultaneously zero, such that
	\begin{align}
		\label{eq: optimization condition}
			& \inprod{\adjCost \derivative{\process}{\proCost(\optProc)} + \sum_{t=0}^{\horizon-1} \dualGroup  \derivative{\process}{\proAlgDyn (\optProc)} + \sum_{t=0}^{\horizon-1} \adjDyn \derivative{\process}{\proStateDyn (\optProc)}}{\perturb{\process}} \nonumber \\ 
			& + \inprod{\sum_{t=0}^{\horizon-1} \adjAuxState \derivative{\process}{\proAuxDyn (\optProc)} + \sum_{t=1}^{\horizon} \adjState \derivative{\process}{\proStCon (\optProc)}}{\perturb{\process}} \le 0\\
			& \quad \quad \text{for } \perturb{\process} + \optProc \in \Bigl(\bigcap_{t=0}^{\horizon-1} Q_{\proCon}(\optProc) \Bigr) \cap Q_{\constrSet{0}{}} (\optProc) \cap Q_{\constrSet{\horizon}{}} (\optProc), \nonumber
	\end{align}
	where \(Q_{K}(r)\) represents a \emph{tent} of the set \(K\) at \(r \in K\) as defined in \cite[\S3]{ref:Bol-75},\footnote{	Let \(\vertex \in \OMG \subset \R^{\genDim}\). A convex cone \(Q_{} \subset \R^{\genDim}\) with vertex \(\vertex\) is a \embf{tent} of \(\OMG\) at \(\vertex\) if there exists a smooth map \(\tentMap\) defined in a neighbourhood \ball[]{\vertex} of \(\vertex\) such that: \(\tentMap (\genVar) = \genVar + o  (\genVar - \vertex )\), and there exists \(\epsilon > 0\) such that \(\tentMap (\genVar) \in \OMG\) for \(\genVar \in Q \cap \ball{\vertex}\). Recall the Landau notation \(\genFun (\genVar) = o (\genVar)\) that stands for a function \(\genFun(0) = 0\) and \(\lim_{\genVar \to 0} \frac{\abs{\genFun(\genVar)}}{\abs{\genVar}} = 0\).}
	and for each \(j \in \scaleZ[\stConDim+1]\), \({(\adjState)}^{(j)} \leq 0 \) and \({(\adjState)}^{(j)} \proStCon^{(j)} (\optProc) = 0 \text{ for each } t \in \scaleZ[\horizon + 1]\).

		\subsubsection*{\ref{pf:step4} First order necessary conditions in the configuration space}
	We push back the optimality condition \eqref{eq: optimization condition} to the configuration manifold \(\prodManifold\). Let \(\adjGroup \Let \algToEucl^{\ast}(\dualGroup)\).  

	\embf{Adjoints of the states \(\state\)}: For each \(t \in \scaleZ\), if we choose \(\perturb{\process}\) in \eqref{eq: optimization condition} such that all its entries except those in \(\proj{\state} (\perturb{\process})\) are zero, then
	\begin{equation}
		\label{e:state adjoints}
		\begin{aligned}
 			& \inprod{\adjCost \derivative{\state} \cost (\optConfig, \optState, \optInp) + \adjState \derivative{\state} \stateConstr(\optConfig, \optState, \optInp)}{\proj{\state} (\perturb{\process})} \\
 			& + \inprod{\adjGroup \derivative{\state} (\exp \inverse \circ \intStep[t]) (\optConfig, \optState, \optInp)}{\proj{\state} (\perturb{\process})} \\
 			& +  \inprod{\adjDyn \derivative{\state} \sysDyn (\optConfig, \optState, \optInp) - \adjDyn[t-1]}{\proj{\state} (\perturb{\process})} \le 0.
		\end{aligned}
	\end{equation}
	Since \(\proj{\state} (\perturb{\process}) \in \R^{\sysDim}\) can be arbitrary, in view of the Hamiltonian defined in \eqref{e:Hamiltonian}, the inequality \eqref{e:state adjoints} reduces to the dynamics of the adjoints \(\adjDyn[]\) in \ref{pmp:dynamics}. An identical procedure can be adopted to derive the transversality condition for \(\adjDyn[\horizon-1]\) in \ref{pmp:transversality} after setting all entries of \(\perturb{\process}\) in \eqref{eq: optimization condition} to zero except \(\proj[\horizon]{\state} (\perturb{\process})\).

	\embf{Adjoints of the (auxiliary) states \(\auxState[]\)}: For each \(t \in \scaleZ[\horizon+1]\), we choose \(\perturb{\process}\) in \eqref{eq: optimization condition} such that all its entries except those in \(\proj{\auxState[]} (\perturb{\process})\) are zero. The optimality condition \eqref{eq: optimization condition} then leads to \(\inprod{\adjAuxState[t-1] - \adjAuxState}{\proj{\auxState[]}(\perturb{\process})} \le 0.\)
	Since \(\proj{\auxState[]}(\perturb{\process}) \in \R^{\freqDim}\) can be picked arbitrarily, \(\adjAuxState[t-1] = \adjAuxState\) for each \(t \in \scaleZ[\horizon]\).
	We define \(\R^{\freqDim} \ni \adjFreq \Let \adjAuxState[0] = \cdots = \adjAuxState[\horizon-1]\); this is our `multiplier' corresponding to the frequency constraints.

	\embf{Adjoints of the configurations \(\config\):}
		Let 
			\begin{equation*}				
				\perturb{\config}_{t} \Let \derivative{\EuclConfigSeq}	\EuclToGroup(\proj[]{\EuclConfig} (\optProc)) 	\proj[]{\EuclConfig} (\perturb{\process}) \in  \tanLift[\optConfig]{ \LieGroup } \quad \text{for } t \in \Scale[N+1],
 			\end{equation*} (cf. \eqref{eq: optimization condition}) be the velocity vector at the configuration $\optConfig$ and \(\chi_{t} \Let T_{\optConfig} \Phi_{{\optConfig}^{-1}} (\perturb{\config}_{t}) \in \LieAlg\) be its correspoding vector in the Lie algebra $ \LieAlg$. Therefore  
			\begin{equation}
				\label{eq:vecE2G}
				T_{e}\Phi_{\optConfig}(\chi_t) = \derivative{\EuclConfigSeq}	\EuclToGroup(\proj[]{\EuclConfig} (\optProc)) 	\proj[]{\EuclConfig} (\perturb{\process}) \in  \tanLift[\optConfig]{ \LieGroup }, t \in \Scale[N+1],
 			\end{equation}
and \(\proj[]{\EuclConfig} (\perturb{\process})\) can be obtained via the tangent lift of \(\proTransfer \inverse\) with \(\kappa_{t-1} \Let \dexp \) as \(\proj[0]{\EuclConfig} (\perturb{\process}) \Let \algToEucl (\chi_{0})\) and 
 			\begin{equation}
				\label{eq:vecG2E} 			 	
				\proj{\EuclConfig} (\perturb{\process}) \Let - \algToEucl \circ \kappa_{t-1}\bigl(\adjAction_{{\optConfig} \inverse \optConfig[t-1]} (\chi_{t-1}) - \chi_{t} \bigr) \quad \text{for } t \in \scaleZ[\horizon+1].
 			\end{equation}
			If we choose \(\perturb{\process}\) in \eqref{eq: optimization condition} such that all its entries except \(\proj[]{\EuclConfig} (\perturb{\process})\) are zero, then by substituting \(\proj{\EuclConfig} (\perturb{\process})\) from \eqref{eq:vecG2E}  and using \eqref{eq:vecE2G}, \eqref{eq: optimization condition} reduces to
			\begin{gather}
				\label{e:config adjoints}
 				\begin{aligned}
 					& \sum_{t=0}^{\horizon-1} \inprod{\adjCost \derivative{\config} \cost(\optConfig, \optState, \optInp) + \adjDyn \derivative{\config} \sysDyn(\optConfig, \optState, \optInp)}{T_{e}\Phi_{\optConfig}(\chi_t)} \\
 					& + \sum_{t=1}^{\horizon} \inprod{\adjState \derivative{\config} \stateConstr(\optConfig, \optState, \optInp)}{T_{e}\Phi_{\optConfig}(\chi_t)} \\
 					& + \sum_{t=0}^{\horizon-1} \inprod{\adjGroup}{\kappa_{t-1}\bigl(\adjAction_{{\optConfig[t+1]}  \inverse \optConfig[t]} (\chi_{t}) - \chi_{t+1}\bigr)} \le 0.
 				\end{aligned} \raisetag{3\baselineskip}
 			\end{gather}
			Since \(\chi_t \in \LieAlg\) can be selected arbitrarily for all \(t \in \Scale[N+1]\), we choose a sequence \((\chi_{\tau})_{\tau=0}^{\horizon}\) such that \(\chi_{\tau}= 0\) for all \(\tau \in \Scale[\horizon +1],\  \tau \neq t\). In view of the above and by definition of the Hamiltonian \eqref{e:Hamiltonian}, \eqref{e:config adjoints} reduces to the dynamics of \(\adjGroup[]\) in \ref{pmp:dynamics} and the transversality condition of \(\adjGroup[\horizon-1]\) in \ref{pmp:transversality} of Theorem \ref{th:main}.

 	\embf{Conditions on the control}:
		Note that \(\admControl\) is convex, and therefore, \(\admControl \subset Q_{\admControl} (\optInp)\). We restrict \(\perturb{\process}\) such that \(\perturb{\conInp_{t}}+\optInp \in Q_{\admControl}(\optInp)\), which simplifies the condition \eqref{eq: optimization condition} to
 		\begin{align*}
 			& \inprod{\adjCost \derivative{\conInp}\cost(\optConfig,\optState,\optInp) + \adjDyn \derivative{\conInp} \sysDyn( \optConfig, \optState, \optInp)}{\omega - \optInp}\\
 			& \quad \quad + \inprod{\adjFreq \freqDer}{\omega-\optInp}\leq 0 \quad \text{for all } \omega \in \admControl .
 		\end{align*}
 		By definition of the Hamiltonian \eqref{e:Hamiltonian}, the preceding condition is equivalent to \ref{pmp:ham maximization} of Theorem \ref{th:main}.

	\section{Numerical experiments}
		\label{sec:numerical simulations}
		We illustrate our theory in the context of a spacecraft attitude manoeuvre, where the configuration manifold is the matrix Lie group \(\SO(3)\) (the set of \(3 \times 3\) real orthogonal matrices with determinant \(1\)) and its Lie algebra is \( \so(3) \) (the set of \(3 \times 3\) real skew-symmetric matrices).

\subsubsection*{Satellite attitude dynamics}

Let \m{\ornt{t}} and \m{\rot{t} \in \OrntMan{3}} be the rotation matrices that relate coordinates 
of a point in the spacecraft body frame to the inertial frame and the change in the orientation at \m{t}th time instant respectively. Let \m{\mtm{t} \in \R^3} be the spacecraft momentum vector in the body frame, and \m{\conInp_{t} \in \R^3} be the torque applied to the spacecraft in the body frame. Assuming that a constant control is applied between two discrete-time instants for a step length \m{h>0} (selected so that the conditions \ref{diffeo} and \ref{uniformity} in \secref{sec:the prob} are met), the discrete-time attitude dynamics in this setting is given in the spacecraft body frame in a standard way as
\begin{align}\label{eq:AttitudeModel}
\begin{cases} 
\ornt{t+1} \nsp&=\ornt{t}\rot{t},\\ 
\mtm{t+1} \nsp&=\rot{t}^\top \mtm{t} + h \conInp_{t},\\ \widehat{h\mtm{t}} \nsp&= \rot{t}\mint - \mint\rot{t}^\top,
\end{cases}
\end{align}
where \m{\mint \Let \tr(\mathcal{J})I-\mathcal{J}} and \m{\mathcal{J}} is the moment of inertia matrix. A detailed derivation of the system dynamics is given in, e.g., \cite{MSM15}. The data for our numerical experiments are:
\begin{table}[h]
  \centering
  \begin{tabular}{l|l}
  \hline 
  \textbf{System Parameters and Specifications} & \textbf{ Admissible range} \\ \hline
    the principal moment of inertia of  & \\
    the spacecraft \(\left(\Inertia{x},\Inertia{y},\Inertia{z}\right)\)  & \(\left(800,1200,1000\right)  \si{\kilo\gram\square\meter}\) \\
	the sampling time ($T$) &  0.1 \si{\second} \\
	  manoeuvre angle range ($\theta$) (any axis) & [10 \si{\degree}, 90 \si{\degree}]\\ 
	the maximum admissible torque & \\ 
     bound (control bound) ($u_{\text{bnd}}$) &  \(\left(20,20,20\right) \si{\newton\meter}\) \\
	the maximum momentum ($ \Pi_{\text{bnd}}$) & \(\left(60, 60, 60\right) \si{\newton\meter\second}\) \\ 
	frequency constraints on the control \( \conInp\kth[1], \conInp\kth[3]\)  & $0$ above $2 \pi/3$ \\
	frequency constraints on the control - $\conInp\kth[2]$  &   no restriction \\
	range of the time duration ($t_{\max}$) of & \\ manoeuvres & 5 \si{\second} and 30 \si{\second} \\ \hline 
  \end{tabular} 
  \label{tab:table1}
\end{table}

\subsubsection*{Problem definition}: 
\label{ssec:PD}
The control objective is to synthesize an energy-optimal control profile to manoeuvre a spacecraft from the initial state \m{(\ornt{i},\mtm{i})} to the final state \m{(\ornt{f},\mtm{f})} in \m{N} discrete-time instances while satisfying the following state and control constraints:
\begin{equation}\label{eq:InqConst}
\begin{aligned}
\begin{cases}	
	\abs{\conInp_{t}^{(i)}}\leq \conInp_{\text{bnd}}^{(i)} \quad & \text{for } t \in \Scale\\
	\abs{\mtm{t}^{(i)}}\leq\mtm{\text{bnd}}^{(i)} \; & \text{for } t \in \scaleZ[\horizon] \\
	\support(\widehat{u^{(i)}}) \subset \mathcal{W}_i
\end{cases} \quad \text{for} \quad i=1,2,3,
\end{aligned}
\end{equation}
where \( \mathcal{W}_{i} \Let \set{i \suchthat \frac{2\pi (i-1)}{\horizon} \leq \frac{2 \pi}{3} \text{ or } \frac{2\pi (i-1)}{\horizon}\geq \frac{4 \pi }{3}}\) for \(i = 1, 3\) and \(  \mathcal{W}_{2} =  \set{1, \ldots, \horizon} \) denote the set of allowable frequencies in the torque profile \m{u^{(i)}}, \m{\conInp_{\text{bnd}}^{(i)} \in \R^{+}} denotes the torque bound on the actuator along \m{i}-th axis and \m{\mtm{\text{bnd}}^{(i)} \in \R^{+} } denotes the momentum bound along \m{i}-th axis.

Our frequency-constrained optimal control problem is:
\begin{equation}\label{eq:opt}
	\begin{aligned}
		& \minimize_{\{\conInp_{t}\}_{t=0}^{N-1}} && \mathcal{J}(u)\Let\sum_{t=0}^{N-1} \frac{1}{2} \norm{\conInp_{t}}^{2}_{2} \\
		& \sbjto && \text{dynamics}\; \eqref{eq:AttitudeModel} \text{ and constraints } \eqref{eq:InqConst}.
	\end{aligned}
\end{equation}
  
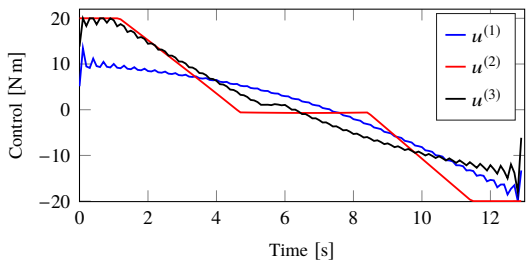
\begin{figure}[h!]
	\centering	
	\setlength\figureheight{=0.18\textwidth} 
	\setlength\figurewidth{=0.35\textwidth}	
%
%
\begin{tikzpicture}

\begin{axis}[%
width=0.951\figurewidth,
height=0.8\figureheight,
at={(0\figurewidth,0\figureheight)},
scale only axis,
xmin=0,
xmax=13,
xlabel={Time $[\si{\second}]$},
ymin=-20.1,
ymax=22,
ylabel={Control $[\si{\newton\meter}]$},
y label style={at={(axis description cs:-0.1,.5)},rotate=0,anchor=south},
label style={font=\scriptsize},
tick label style={font=\scriptsize},
axis background/.style={fill=white},
title style={font=\bfseries},
legend image post style={scale=0.45},
legend style={font=\scriptsize,legend cell align=left,align=left,draw=white!15!black}
]
\addplot [color=blue,solid,line width=0.7pt]
  table[row sep=crcr]{%
0	5.1391660387155\\
0.1	13.480095460449\\
0.2	9.47202803004429\\
0.3	9.13738483884369\\
0.4	11.1951682068617\\
0.5	9.53878030602925\\
0.6	9.33609837169863\\
0.7	10.4740280700239\\
0.8	9.39356869141645\\
0.9	9.24610753616588\\
1	10.0135448891472\\
1.1	9.19080197081727\\
1.2	9.0731713717477\\
1.3	9.6401540245885\\
1.4	8.96204055586068\\
1.5	8.86240320088538\\
1.6	9.30308439403001\\
1.7	8.71605327201718\\
1.8	8.62760903835312\\
1.9	8.98066903143562\\
2	8.45494714899833\\
2.1	8.3733512904039\\
2.2	8.66136507595541\\
2.3	8.17853060073512\\
2.4	8.10076265177108\\
2.5	8.33793857966875\\
2.6	7.8856805492928\\
2.7	7.80951141432544\\
2.8	8.00529379052371\\
2.9	7.57489087441716\\
3	7.49860516224948\\
3.1	7.65953755149857\\
3.2	7.24452200972912\\
3.3	7.16676219572093\\
3.4	7.29753064414461\\
3.5	6.89292611033082\\
3.6	6.81260000343435\\
3.7	6.91665460549006\\
3.8	6.51851496054665\\
3.9	6.43473827266919\\
4	6.51468287798388\\
4.1	6.11979932036835\\
4.2	6.03185839802925\\
4.3	6.08970696907245\\
4.4	5.69541242996058\\
4.5	5.60273704094724\\
4.6	5.64010489336355\\
4.7	5.24401430767492\\
4.8	5.14530425412185\\
4.9	5.16152386519495\\
5	4.75926191185643\\
5.1	4.65238891559132\\
5.2	4.64780925549251\\
5.3	4.23768909520438\\
5.4	4.12287880249969\\
5.5	4.09967453497607\\
5.6	3.68263769889915\\
5.7	3.56248757963154\\
5.8	3.52293031704217\\
5.9	3.09764323721758\\
6	2.97135232872555\\
6.1	2.91454965767949\\
6.2	2.47735965695869\\
6.3	2.34279512834067\\
6.4	2.26829303901837\\
6.5	1.81790493716402\\
6.6	1.67539288259941\\
6.7	1.58380462513318\\
6.8	1.11911539469688\\
6.9	0.969086159812054\\
7	0.861066515182337\\
7.1	0.381036701721105\\
7.2	0.22408364201763\\
7.3	0.100330768420522\\
7.4	-0.396049699289215\\
7.5	-0.559150467836361\\
7.6	-0.697884670036638\\
7.7	-1.21163185248894\\
7.8	-1.37990117520782\\
7.9	-1.53279934395029\\
8	-2.06497777900243\\
8.1	-2.23720679078238\\
8.2	-2.40336987064652\\
8.3	-2.95516207007913\\
8.4	-3.12983144546558\\
8.5	-3.30769426012423\\
8.6	-3.8788687528437\\
8.7	-4.05202622530559\\
8.8	-4.23868972912083\\
8.9	-4.82995100755611\\
9	-4.99930929094252\\
9.1	-5.19365168080755\\
9.2	-5.80698940223516\\
9.3	-5.96999818267907\\
9.4	-6.17076362782697\\
9.5	-6.80870999890709\\
9.6	-6.96223932574544\\
9.7	-7.16800876250385\\
9.8	-7.83386955040567\\
9.9	-7.97400231645661\\
10	-8.18313642870192\\
10.1	-8.88133844067815\\
10.2	-9.00305440827289\\
10.3	-9.21360090627533\\
10.4	-9.95023814666434\\
10.5	-10.0469091158411\\
10.6	-10.2564474266505\\
10.7	-11.0401820813312\\
10.8	-11.1027255845418\\
10.9	-11.3080913177864\\
11	-12.1517308448004\\
11.1	-12.1671047562637\\
11.2	-12.3638624927723\\
11.3	-13.2873337254649\\
11.4	-13.2356476160918\\
11.5	-13.4169755487127\\
11.6	-14.4535209056442\\
11.7	-14.3019741904724\\
11.8	-14.4560843929778\\
11.9	-15.6671507064392\\
12	-15.3552876880643\\
12.1	-15.4572013996227\\
12.2	-16.974974392478\\
12.3	-16.3704446103819\\
12.4	-16.3452935569483\\
12.5	-18.5470574361092\\
12.6	-17.2848472673089\\
12.7	-16.523103104718\\
12.8	-20\\
12.9	-13.2847813388084\\
};
\addlegendentry{${u}^{(1)}$};

\addplot [color=red,solid,line width=0.7pt]
  table[row sep=crcr]{%
0	20\\
0.1	20\\
0.2	20\\
0.3	20\\
0.4	20\\
0.5	20\\
0.6	20\\
0.7	20\\
0.8	20\\
0.9	20\\
1	20\\
1.1	20\\
1.2	19.786546945256\\
1.3	19.2153633279514\\
1.4	18.6439851551654\\
1.5	18.0723238364501\\
1.6	17.500313654703\\
1.7	16.9279369224217\\
1.8	16.3551162925684\\
1.9	15.7817924379501\\
2	15.2079427766735\\
2.1	14.6334990361415\\
2.2	14.0584076386606\\
2.3	13.482643916521\\
2.4	12.9061475709006\\
2.5	12.3288704958832\\
2.6	11.7507877816756\\
2.7	11.1718465056317\\
2.8	10.5920039447179\\
2.9	10.0112362535054\\
3	9.42949755584493\\
3.1	8.84675052028885\\
3.2	8.26297333773577\\
3.3	7.67812699697776\\
3.4	7.09217961242595\\
3.5	6.50511215426756\\
3.6	5.91689237390817\\
3.7	5.32749389892523\\
3.8	4.73690105076389\\
3.9	4.14508827713065\\
4	3.55203477729084\\
4.1	2.95772865776148\\
4.2	2.36215100341674\\
4.3	1.76528661988163\\
4.4	1.16712771282172\\
4.5	0.567661932344974\\
4.6	-0.0331203083529767\\
4.7	-0.583262558648304\\
4.8	-0.594148709773667\\
4.9	-0.604846497699903\\
5	-0.615231385590007\\
5.1	-0.624374526452879\\
5.2	-0.633264291277949\\
5.3	-0.641775650975596\\
5.4	-0.649022975160093\\
5.5	-0.656094387585715\\
5.6	-0.663130916889539\\
5.7	-0.669429536636046\\
5.8	-0.675774712343142\\
5.9	-0.681947347608908\\
6	-0.687289711720523\\
6.1	-0.692524498909739\\
6.2	-0.697084478516095\\
6.3	-0.700332907556968\\
6.4	-0.703361264427667\\
6.5	-0.705733143326425\\
6.6	-0.706761460866322\\
6.7	-0.707550850504762\\
6.8	-0.707671060744164\\
6.9	-0.706435837899112\\
7	-0.704964034135078\\
7.1	-0.70281311869983\\
7.2	-0.699309870966091\\
7.3	-0.695588081538456\\
7.4	-0.691185958140929\\
7.5	-0.685448466842542\\
7.6	-0.679524767080266\\
7.7	-0.672930212988673\\
7.8	-0.665030744163289\\
7.9	-0.656991131174693\\
8	-0.648301650284356\\
8.1	-0.638350954034097\\
8.2	-0.628319542345528\\
8.3	-0.617671044604526\\
8.4	-0.605817961650896\\
8.5	-1.12005620324246\\
8.6	-1.75753837289836\\
8.7	-2.39542977296538\\
8.8	-3.0336961975282\\
8.9	-3.67230724066799\\
9	-4.31122182270829\\
9.1	-4.95040672618305\\
9.2	-5.58983330802395\\
9.3	-6.22946088712469\\
9.4	-6.86925758683838\\
9.5	-7.50919693985791\\
9.6	-8.14923889216841\\
9.7	-8.7893532022339\\
9.8	-9.4295159602686\\
9.9	-10.0696878712884\\
10	-10.7098405560129\\
10.1	-11.3499530135153\\
10.2	-11.9899867214964\\
10.3	-12.6299153142373\\
10.4	-13.2697210274087\\
10.5	-13.909365978397\\
10.6	-14.5488258823273\\
10.7	-15.1880865326819\\
10.8	-15.8271103666068\\
10.9	-16.4658751556969\\
11	-17.1043706014782\\
11.1	-17.7425588912242\\
11.2	-18.3804197297819\\
11.3	-19.0179471776542\\
11.4	-19.6551022444125\\
11.5	-20\\
11.6	-20\\
11.7	-20\\
11.8	-20\\
11.9	-20\\
12	-20\\
12.1	-20\\
12.2	-20\\
12.3	-20\\
12.4	-20\\
12.5	-20\\
12.6	-20\\
12.7	-20\\
12.8	-20\\
12.9	-20\\
};
\addlegendentry{${u}^{(2)}$};

\addplot [color=black,solid,line width=0.7pt]
  table[row sep=crcr]{%
0	13.8030255064746\\
0.1	20\\
0.2	18.3317427725094\\
0.3	20\\
0.4	20\\
0.5	18.6628137749329\\
0.6	20\\
0.7	20\\
0.8	19.04241954712\\
0.9	20\\
1	19.5734203321331\\
1.1	18.2471498041249\\
1.2	18.728151715314\\
1.3	18.2609957461452\\
1.4	16.9960360232651\\
1.5	17.287663987469\\
1.6	16.8652699077867\\
1.7	15.7206818425558\\
1.8	15.8901471994182\\
1.9	15.4775114353778\\
2	14.4286845244275\\
2.1	14.5172098673313\\
2.2	14.1053228996465\\
2.3	13.131693693489\\
2.4	13.1641715917259\\
2.5	12.7509913995985\\
2.6	11.8378203922977\\
2.7	11.8304613763468\\
2.8	11.4162582145877\\
2.9	10.5530360033917\\
3	10.5170739832142\\
3.1	10.1029643751993\\
3.2	9.28207443265236\\
3.3	9.22567164103253\\
3.4	8.81310356406594\\
3.5	8.02891340544357\\
3.6	7.95819805330251\\
3.7	7.54878202971156\\
3.8	6.79703772967273\\
3.9	6.71667587306521\\
4	6.31217957695032\\
4.1	5.58959616682923\\
4.2	5.50306036817711\\
4.3	5.10553694844699\\
4.4	4.40952292424496\\
4.5	4.31905764523564\\
4.6	3.9311980030693\\
4.7	3.25966043575404\\
4.8	3.16507574920314\\
4.9	2.78962779181715\\
5	2.140167242123\\
5.1	2.03577801747235\\
5.2	1.67661254023338\\
5.3	1.07476669822891\\
5.4	1.08326789251436\\
5.5	1.09508785471406\\
5.6	1.00543025950265\\
5.7	1.21898762894352\\
5.8	1.12798574636112\\
5.9	0.96976364688324\\
6	1.0523792073756\\
6.1	0.556852821365778\\
6.2	0.0140090064178983\\
6.3	0.0215391443648388\\
6.4	-0.414905831453755\\
6.5	-0.953304197564193\\
6.6	-0.947090725381104\\
6.7	-1.36241863184187\\
6.8	-1.89018181987905\\
6.9	-1.87580939226481\\
7	-2.27853608947249\\
7.1	-2.79449331429458\\
7.2	-2.76765271157391\\
7.3	-3.16145192995823\\
7.4	-3.66570550027607\\
7.5	-3.6234790182557\\
7.6	-4.0105840006322\\
7.7	-4.50375892037531\\
7.8	-4.44371680176518\\
7.9	-4.82587541062471\\
8	-5.30891004510863\\
8.1	-5.22875964783671\\
8.2	-5.60760713127279\\
8.3	-6.08171401058351\\
8.4	-5.97905991603447\\
8.5	-6.35576910372163\\
8.6	-6.82092198823464\\
8.7	-6.69115199400465\\
8.8	-7.0659796899589\\
8.9	-7.5231603292002\\
9	-7.36329524686387\\
9.1	-7.73869061702569\\
9.2	-8.19015582171528\\
9.3	-7.99686699094029\\
9.4	-8.37577024590002\\
9.5	-8.82427995479589\\
9.6	-8.59344869740245\\
9.7	-8.97946205021589\\
9.8	-9.42843950884154\\
9.9	-9.15477347940634\\
10	-9.55238886849055\\
10.1	-10.0061688264293\\
10.2	-9.68262302600604\\
10.3	-10.0975571303461\\
10.4	-10.5617865840351\\
10.5	-10.178642671869\\
10.6	-10.6183594696787\\
10.7	-11.1006769362162\\
10.8	-10.6439996268853\\
10.9	-11.1185714331938\\
11	-11.6298310948349\\
11.1	-11.0787120619573\\
11.2	-11.6023290501357\\
11.3	-12.1589843663748\\
11.4	-11.4802152320283\\
11.5	-12.0740417960991\\
11.6	-12.7032735454875\\
11.7	-11.8398830195694\\
11.8	-12.537968415776\\
11.9	-13.290303754971\\
12	-12.1330771743915\\
12.1	-12.996729087557\\
12.2	-13.9846695867719\\
12.3	-12.2838486639361\\
12.4	-13.4444116078433\\
12.5	-15.0072220501397\\
12.6	-11.9615959048441\\
12.7	-13.8058755654928\\
12.8	-18.0367033861825\\
12.9	-6.14138213757983\\
};
\addlegendentry{${u}^{(3)}$};

\end{axis}
\end{tikzpicture}%
	\caption{Optimal control profiles.}
\label{fig:OptimalControl}	
\end{figure}

\begin{figure}[h!]
	\centering	
	\setlength\figureheight{=0.30\textwidth} 
	\setlength\figurewidth{=0.35\textwidth}	
	\input{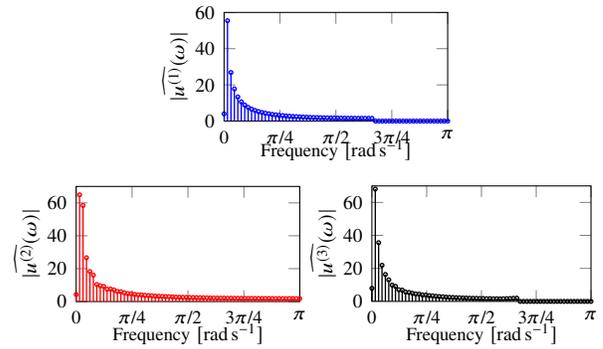}	
	\caption{Frequency spectra of the optimal controls.}
	\label{fig:Frequency_Profile}	
\end{figure}

\subsubsection*{Trajectory generation}
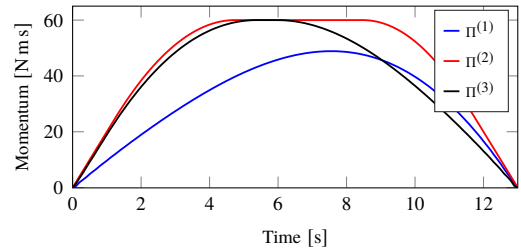
\begin{figure}[h!]
	\centering	
	\setlength\figureheight{=0.17\textwidth} 
	\setlength\figurewidth{=0.35\textwidth}	
%
%
\begin{tikzpicture}

\begin{axis}[%
width=0.951\figurewidth,
height=0.8\figureheight,
at={(0\figurewidth,0\figureheight)},
scale only axis,
xmin=0,
xmax=13,
xlabel={Time $[\si{\second}]$},
ymin=-0.1,
ymax=65,
ylabel={Momentum $[\si{\newton\meter\second}]$},
y label style={at={(axis description cs:-0.07,.5)},rotate=0,anchor=south},
axis background/.style={fill=white},
title style={font=\bfseries},
label style={font=\scriptsize},
tick label style={font=\scriptsize},
legend image post style={scale=0.35},
legend style={font=\tiny,legend cell align=left,align=left,draw=white!15!black}
]
\addplot [color=blue,solid,line width=0.7pt]
  table[row sep=crcr]{%
0	0\\
0.1	0.513926099906207\\
0.2	1.86199313330318\\
0.3	2.80943194781789\\
0.4	3.7237034978636\\
0.5	4.84419449007915\\
0.6	5.79962016863204\\
0.7	6.73545851689999\\
0.8	7.78592587453498\\
0.9	8.72931589224009\\
1	9.65903328812238\\
1.1	10.6667261848532\\
1.2	11.5934933444109\\
1.3	12.5099231428237\\
1.4	13.4846096374231\\
1.5	14.3931185920815\\
1.6	15.2933242199649\\
1.7	16.2393496789751\\
1.8	17.1284728778098\\
1.9	18.0105513356532\\
2	18.9297999687667\\
2.1	19.7983646283504\\
2.2	20.6606323573889\\
2.3	21.5536042829996\\
2.4	22.4001964728624\\
2.5	23.2408685113302\\
2.6	24.1071331110231\\
2.7	24.9300268794649\\
2.8	25.7470928425786\\
2.9	26.5855251308123\\
3	27.3826667214661\\
3.1	28.1738473115385\\
3.2	28.9827708274862\\
3.3	29.7517879496091\\
3.4	30.5145286538739\\
3.5	31.2918134808526\\
3.6	32.03003706908\\
3.7	32.7615206348832\\
3.8	33.504657985397\\
3.9	34.2091506021213\\
4	34.9063196228765\\
4.1	35.612483301778\\
4.2	36.2800711996273\\
4.3	36.9396563345454\\
4.4	37.6057564502481\\
4.5	38.2330635764345\\
4.6	38.8516146793118\\
4.7	39.4743476809387\\
4.8	40.0578194628541\\
4.9	40.6316468786195\\
5	41.2073120044604\\
5.1	41.7429276991758\\
5.2	42.2679666864187\\
5.3	42.7926465715362\\
5.4	43.2763760140306\\
5.5	43.7486244952801\\
5.6	44.2185521038489\\
5.7	44.6467755330341\\
5.8	45.0629734168285\\
5.9	45.4752252112851\\
6	45.8449477426723\\
6.1	46.2020239254443\\
6.2	46.5534098837757\\
6.3	46.8610164395346\\
6.4	47.1550511398062\\
6.5	47.4415201752434\\
6.6	47.6827905695988\\
6.7	47.9095955376761\\
6.8	48.1270274396357\\
6.9	48.2977340864077\\
7	48.4531283262175\\
7.1	48.5974120885941\\
7.2	48.6933437059896\\
7.3	48.7731790679259\\
7.4	48.8402405929503\\
7.5	48.8572259541404\\
7.6	48.8574127006416\\
7.7	48.8432416185035\\
7.8	48.7771727114547\\
7.9	48.6937046569988\\
8	48.5943806804804\\
8.1	48.4412347611764\\
8.2	48.2702139605968\\
8.3	48.0819331995435\\
8.4	47.8377921241803\\
8.5	47.5754562454525\\
8.6	47.2946169914078\\
8.7	46.9558631710946\\
8.8	46.598900088568\\
8.9	46.2223426117616\\
9	45.7856491482759\\
9.1	45.3309234802207\\
9.2	44.8556444218755\\
9.3	44.3178403844337\\
9.4	43.7624688509458\\
9.5	43.1857441466833\\
9.6	42.5438868713035\\
9.7	41.8852754484764\\
9.8	41.2046894379594\\
9.9	40.4560699574015\\
10	39.6919394990501\\
10.1	38.9054157042393\\
10.2	38.047554255912\\
10.3	37.1759671028995\\
10.4	36.2818036319552\\
10.5	35.312430001505\\
10.6	34.3318203986798\\
10.7	33.3287355051563\\
10.8	32.2457441421078\\
10.9	31.1549576839674\\
11	30.0421564854888\\
11.1	28.8435152889191\\
11.2	27.6418680777152\\
11.3	26.4191542795344\\
11.4	25.1027188296977\\
11.5	23.79010056675\\
11.6	22.4580910678331\\
11.7	21.0212085311461\\
11.8	19.5983182601076\\
11.9	18.1589753154865\\
12	16.5975394137003\\
12.1	15.0663640373422\\
12.2	13.524195317226\\
12.3	11.8295064378873\\
12.4	10.1945927810307\\
12.5	8.56164310907775\\
12.6	6.70802934271415\\
12.7	4.98021934667169\\
12.8	3.32829386100349\\
12.9	1.32845507474185\\
13	0\\
};
\addlegendentry{$\Pi^{(1)}$};

\addplot [color=red,solid,line width=0.7pt]
  table[row sep=crcr]{%
0	0\\
0.1	2\\
0.2	4.0000177346549\\
0.3	6.00017508674249\\
0.4	8.00054124163968\\
0.5	10.0012126905582\\
0.6	12.0023282601505\\
0.7	14.0039341837212\\
0.8	16.0061355831334\\
0.9	18.009068904683\\
1	20.012772159979\\
1.1	22.0173513248684\\
1.2	24.022928146283\\
1.3	26.0082448399161\\
1.4	27.9375424200738\\
1.5	29.8109087460368\\
1.6	31.6283107911085\\
1.7	33.3897942512755\\
1.8	35.0954174012134\\
1.9	36.7451174022172\\
2	38.3389132039512\\
2.1	39.8768339571387\\
2.2	41.3587891410047\\
2.3	42.7847723611949\\
2.4	44.1547847631787\\
2.5	45.4687099585929\\
2.6	46.7265177127778\\
2.7	47.9281825879194\\
2.8	49.0735641114012\\
2.9	50.162609829306\\
3	51.1952693578518\\
3.1	52.1713799880206\\
3.2	53.090868801799\\
3.3	53.9536622961796\\
3.4	54.7595774818291\\
3.5	55.5085228750981\\
3.6	56.200403861908\\
3.7	56.8350192577062\\
3.8	57.4122610555436\\
3.9	57.9320157168956\\
4	58.3940660829187\\
4.1	58.7982898000457\\
4.2	59.1445567695466\\
4.3	59.4326362115263\\
4.4	59.6623937133927\\
4.5	59.833685162457\\
4.6	59.9462686318763\\
4.7	60\\
4.8	60\\
4.9	60\\
5	60\\
5.1	60\\
5.2	60\\
5.3	60\\
5.4	60\\
5.5	60\\
5.6	60\\
5.7	60\\
5.8	60\\
5.9	60\\
6	60\\
6.1	60\\
6.2	60\\
6.3	60\\
6.4	60\\
6.5	60\\
6.6	60\\
6.7	60\\
6.8	60\\
6.9	60\\
7	60\\
7.1	60\\
7.2	60\\
7.3	60\\
7.4	60\\
7.5	60\\
7.6	60\\
7.7	60\\
7.8	60\\
7.9	60\\
8	60\\
8.1	60\\
8.2	60\\
8.3	60\\
8.4	60\\
8.5	60\\
8.6	59.9474169286516\\
8.7	59.8298405755171\\
8.8	59.6471169596481\\
8.9	59.3992169586698\\
9	59.0860553964403\\
9.1	58.7074977269716\\
9.2	58.2635356650729\\
9.3	57.7541016449986\\
9.4	57.1790812252486\\
9.5	56.5384875595116\\
9.6	55.8322711405481\\
9.7	55.0603380722289\\
9.8	54.2227233297402\\
9.9	53.3193957045557\\
10	52.3502820790669\\
10.1	51.3154394385888\\
10.2	50.2148549309387\\
10.3	49.0484762734389\\
10.4	47.8163824903568\\
10.5	46.5185789971273\\
10.6	45.1550342671355\\
10.7	43.7258492763465\\
10.8	42.2310475195296\\
10.9	40.6706180645005\\
11	39.0446836883909\\
11.1	37.3532857344639\\
11.2	35.5964336851129\\
11.3	33.7742719652474\\
11.4	31.8868595595857\\
11.5	29.9342262461601\\
11.6	27.9457539443408\\
11.7	25.9559423129845\\
11.8	23.9647980944252\\
11.9	21.9724637397328\\
12	19.9789887920031\\
12.1	17.9843948154212\\
12.2	15.9888402481918\\
12.3	13.9923876848261\\
12.4	11.9950745759962\\
12.5	9.99707545813535\\
12.6	7.99846710524371\\
12.7	5.99930526596787\\
12.8	3.99977837631944\\
12.9	1.99997959979104\\
13	0\\
};
\addlegendentry{$\Pi^{(2)}$};

\addplot [color=black,solid,line width=0.7pt]
  table[row sep=crcr]{%
0	0\\
0.1	1.38028685435052\\
0.2	3.38024402506805\\
0.3	5.21308999361742\\
0.4	7.21238752842357\\
0.5	9.21114599719587\\
0.6	11.0753882213913\\
0.7	13.0724870986619\\
0.8	15.0685557743063\\
0.9	16.9675821354247\\
1	18.9610292553126\\
1.1	20.9102906191956\\
1.2	22.7251161335278\\
1.3	24.586226934618\\
1.4	26.3987061916435\\
1.5	28.0825234068064\\
1.6	29.7933188093734\\
1.7	31.459623205911\\
1.8	33.0090136963089\\
1.9	34.5728926768181\\
2	36.0929993658293\\
2.1	37.5055462413055\\
2.2	38.9242862095587\\
2.3	40.2991449756005\\
2.4	41.5738108014466\\
2.5	42.8489338601155\\
2.6	44.0799337651595\\
2.7	45.2167026939514\\
2.8	46.3498833559821\\
2.9	47.4387957925295\\
3	48.4384568054308\\
3.1	49.4316759354542\\
3.2	50.3806622145805\\
3.3	51.2446827508762\\
3.4	52.1002915948821\\
3.5	52.9119154324689\\
3.6	53.6423626849885\\
3.7	54.3631076537274\\
3.8	55.0403427874548\\
3.9	55.6398303294116\\
4	56.2288563785961\\
4.1	56.7750889681846\\
4.2	57.2467375944437\\
4.3	57.7075753669283\\
4.4	58.1266034078289\\
4.5	58.4740122072134\\
4.6	58.8105449862541\\
4.7	59.1065798936258\\
4.8	59.3338071404039\\
4.9	59.5501215410357\\
5	59.7274711690632\\
5.1	59.8384202833377\\
5.2	59.9375983475632\\
5.3	59.9995716722022\\
5.4	60\\
5.5	60\\
5.6	60\\
5.7	59.9898657932619\\
5.8	60\\
5.9	60\\
6	59.9831569877927\\
6.1	59.9736444952394\\
6.2	59.9136902452938\\
6.3	59.7985775535807\\
6.4	59.6834451013546\\
6.5	59.5239372635101\\
6.6	59.3098756369106\\
6.7	59.095829630372\\
6.8	58.8396884155461\\
6.9	58.5302283149343\\
7	58.2217768444953\\
7.1	57.8726690887642\\
7.2	57.4716051360376\\
7.3	57.0729840592408\\
7.4	56.6347885615709\\
7.5	56.1459997347853\\
7.6	55.6613901182331\\
7.7	55.1380748226624\\
7.8	54.5654767193771\\
7.9	53.999047343571\\
8	53.3946162555347\\
8.1	52.7421289204547\\
8.2	52.0980391124734\\
8.3	51.4164977859523\\
8.4	50.6880150733516\\
8.5	49.9704079965283\\
8.6	49.215791705935\\
8.7	48.415464208322\\
8.8	47.629190158282\\
8.9	46.8066843601648\\
9	45.9398721274784\\
9.1	45.0907253633473\\
9.2	44.2058804174674\\
9.3	43.2778795348823\\
9.4	42.3714549543022\\
9.5	41.4295322996653\\
9.6	40.4452833933987\\
9.7	39.4868829724148\\
9.8	38.4927682486781\\
9.9	37.4567528983699\\
10	36.4513198222057\\
10.1	35.4094336405964\\
10.2	34.3255597940136\\
10.3	33.2776212772937\\
10.4	32.1918281145093\\
10.5	31.0632985334349\\
10.6	29.9769260027126\\
10.7	28.850441874993\\
10.8	27.6795936787821\\
10.9	26.5583965076794\\
11	25.3936965311069\\
11.1	24.1817863167552\\
11.2	23.0289725469793\\
11.3	21.8277008286513\\
11.4	20.5745763601197\\
11.5	19.3931572284184\\
11.6	18.1560460930992\\
11.7	16.8595257117945\\
11.8	15.6527618886562\\
11.9	14.379366948304\\
12	13.0336728105882\\
12.1	11.8065106756139\\
12.2	10.4955253075975\\
12.3	9.08801277667014\\
12.4	7.85269582963419\\
12.5	6.50314424213502\\
12.6	4.9988229888191\\
12.7	3.80039259176527\\
12.8	2.41855638449613\\
12.9	0.614304244678092\\
13	0\\
};
\addlegendentry{$\Pi^{(3)}$};

\end{axis}
\end{tikzpicture}%
	\caption{Optimal momentum profiles.}
\label{fig:Momentum_Profile}	
\end{figure}

We present a manoeuvre for reorienting the spacecraft by \m{50 \si{\degree}} about the axis \m{\left(\frac{1}{\sqrt{3}},\frac{1}{\sqrt{3}},\frac{1}{\sqrt{3}}\right)} in the body frame from an initial momentum \m{\mtm{i}= \left(0,0,0\right) \si{\newton\meter\second}} to a desired final momentum  \m{\mtm{f} = \left(0,0,0\right) \si{\newton\meter\second}} in  \m{13 \si{\second}}. The frequency components of the torque profiles along the \m{x}- and \m{z}-axes actuators that are above \m{2 \pi/3} are set to be zero by an appropriate choice of matrices \(\mathcal F_i\), and the one along the \m{y}-axis is left unconstrained; Figure~\ref{fig:Frequency_Profile} reflects the outcome.  The optimal controls and the corresponding momentum profiles are shown in Figure~\ref{fig:OptimalControl} and Figure~\ref{fig:Momentum_Profile} respectively. Observe that the optimal control along the \m{x}- and \m{y}-axes saturates for the time duration \m{1-2 \si{\second}} in order to execute the pre-specified manoeuvre within the given time interval, and that the optimal control along the \m{y}-axis becomes zero for the time duration \m{6.2-7 \si{\second}} since the momentum constraints along \m{y}-axis are active for that duration.

		\vspace{-0.45cm}

	 \bibliography{references}
	 \bibliographystyle{IEEEtran}

\end{document}